\documentclass[useAMS,usenatbib]{mn2e}
\usepackage[british]{babel}
\usepackage{graphicx}
\usepackage{amsmath}
\usepackage{mathrsfs}
\usepackage{stmaryrd}

\newcommand{\simnot}{\mathord{\sim}}

\begin{document}

\title[Very Fast X-ray Spectral Variability in Cygnus X-1]{Very Fast X-ray Spectral Variability in Cygnus X-1:\\Origin of the Hard and Soft-State Emission Components}
\author[C.~J.~Skipper, I.~M.~M$^c$Hardy and T.~J.~Maccarone]{Chris~J.~Skipper,$^1$, Ian~M.~M$^c$Hardy$^1$ and Thomas~J.~Maccarone$^{1,2}$ \\ \\ $^1$School of Physics and Astronomy, University of Southampton, Southampton, UK SO17 1BJ \\
$^2$Department of Physics, Texas Tech University, Box 41051, Lubbock, TX  79409-1051, USA}
\maketitle

\begin{abstract}
The way in which the X-ray photon index, $\Gamma$, varies as a function of count rate is a strong diagnostic of the emission processes and emission geometry around accreting compact objects. Here we present the results from a study using a new, and simple, method designed to improve sensitivity to the measurement of the variability of $\Gamma$ on very short time-scales.

We have measured $\Gamma$ in $\simnot$2 million spectra, extracted from observations with a variety of different accretion rates and spectral states, on time-scales as short as $\rm 16~ms$ for the high mass X-ray binary Cygnus X-1 (and in a smaller number of spectra for the low mass X-ray binary GX 339-4), and have cross-correlated these measurements with the source count rate. In the soft-state cross-correlation functions (CCFs) we find a positive peak at zero lag, stronger and narrower in the softer observations. Assuming that the X-rays are produced by Compton scattering of soft seed photons by high energy electrons in a corona, these results are consistent with Compton cooling of the corona by seed photons from the inner edge of the accretion disc, the truncation radius of which increases with increasing hardness ratio.

The CCFs produced from the hard-state observations, however, show an anti-correlation which is most easily explained by variation in the energy of the electrons in the corona rather than in variation of the seed photon flux. The hard-state CCFs can be decomposed into a narrow anti-correlation at zero lag, which we tentatively associate with the effects of self-Comptonisation of cyclo-synchrotron seed photons in either a hot, optically thin accretion flow or the base of the jet, and a second, asymmetric component which we suggest is produced as a consequence of a lag between the soft and hard X-ray emission. The lag may be caused by a radial temperature/energy gradient in the Comptonising electrons combined with the inward propagation of accretion rate perturbations.
\end{abstract}


\section{Introduction}

The X-ray spectra of accreting black holes can often be fitted with just a few common components, such as a power-law continuum, an iron K${\alpha}$ fluorescence line at $\rm\simnot6.4~keV$, a reflection component and, if required, a thermal component possibly representing emission from an accretion disc. The power-law component, which takes the form $F_{\rm X}~{\propto}~E^{-\Gamma}$ (where $\Gamma$ is the photon index, ${F_{\rm X}}$ is the photon flux density and \textit{E} is the energy), is usually attributed to Comptonisation of lower energy seed photons by high energy electrons \citep{Sunyaev1980} in either a central corona or the base of a jet and is observed in both the soft and the hard spectral states.

Previous studies of black hole X-ray spectra have established a correlation, within samples of active galactic nuclei (AGN), between the power-law photon index ($\Gamma$) and the source accretion rate (\textit{\.{m}}${\rm_{E}}$) \cite[e.g.][]{Constantin2009,Gu2009,Shemmer2006}. A similar relationship has also been found within individual X-ray binary systems (XRBs) \cite[e.g.][]{Wu2008,Sobolewska2011,Zdziarski2002}. Above a critical accretion rate (\textit{\.{m}}$_{\rm crit}$), typically $\rm0.5-1\%$ of the Eddington limit, \textit{\.{m}}$_{\rm edd}$, $\Gamma$ is found to be positively correlated with the accretion rate (the source is softer when brighter) while below \textit{\.{m}}$_{\rm crit}$ these two properties are anti-correlated (the source is harder when brighter). This switch in the relationship between \textit{\.{m}}$_{\rm E}$ and $\Gamma$ is not the same as the state transition from the hard to the soft state, which usually occurs at a higher accretion rate of around a few percent of the Eddington limit or greater \cite[e.g.][]{Maccarone2003,Gierlinski2006}.

In all Comptonisation models the way in which $\Gamma$ varies with luminosity, or accretion rate, is a major diagnostic of the physical parameters of the emission region, such as the temperature and optical depth of the high energy electrons and the energy and flux of the seed photons. These parameters, in turn, depend upon the origin of the seed photons and Comptonising electrons, the latter of which may be of uniform energy, or may be energy stratified as a function of radius.

In the soft state the accretion flow is believed to take the form of a radiatively efficient, optically thick, geometrically thin disc at all radii, with the seed photon supply being dominated by black body photons from the inner edge of the accretion disc. The origin of the positive correlation between $\Gamma$ and \textit{\.{m}}$_{\rm E}$ in the soft state is therefore often attributed to Compton cooling of the corona by seed photons from the accretion disc. In the hard state the flow at small radii may become a geometrically thick but optically thin and radiatively inefficient accretion flow (RIAF) \citep{Gu2009,Wu2008,Done2006,Narayan2008,Qiao2010,Niedzwiecki2012} which contributes significantly to seed photon supply through synchrotron and bremsstrahlung emission. The anti-correlation found in the hard state at low accretion rates, when the spectrum is expected to be dominated by Comptonisation of these synchrotron and bremsstrahlung photons, may be due to an injection of high energy electrons into the corona producing both an increase in seed photon emission and a harder Compton-scattered spectrum \citep{Qiao2013}. However, under the truncated disc model \citep{Esin1997}, the inner radius of the accretion disc in the hard state moves inwards with increasing luminosity, and the balance of seed photon supply may consequently shift from synchrotron and bremsstrahlung emission from the RIAF at low \textit{\.{m}}$_{\rm E}$ to disc black body emission at high \textit{\.{m}}$_{\rm E}$.

Within individual AGN the positive $\Gamma$-\textit{\.{m}}$_{\rm E}$ correlation is well established on time-scales of months to years \cite[e.g.][]{Lamer2003,Sobolewska2009} as well as on very short time-scales of hours \citep{Ponti2006}. The negative correlation has also been found in one AGN on long time-scales \citep{Emmanoulopoulos2012}. Within individual XRBs both the positive and negative correlations have been observed on long time-scales of days to years \citep{Wu2008,Sobolewska2011,Zdziarski2002}.

Within individual XRBs, however, there is little information available on the relationship between $\Gamma$ and \textit{\.{m}}$_{\rm E}$ on very short time-scales (i.e. comparable, when scaled by mass, with the relatively short timescales on which X-ray spectral variability has been studied in AGN). \cite{Wu2010} showed that the positive correlation exists on time-scales of $\rm 62~ms$ in a single observation of Cygnus X-1, but had to correct for a significant amount of Poisson noise in order to do so. This time-scale corresponds approximately to the Compton cooling time-scale for the corona around a $\rm10~M_{\varodot}$ black hole \citep{Ishibashi2012} or the viscous time-scale and so is one of the shortest time-scales on which observable changes in the emission region may be expected to occur. Scaled by mass, this time-scale also corresponds to approximately $\rm17~hours$ for an AGN with a ${\rm 10^{6}~{M_{\varodot}}}$ black hole and thus allows us to study spectral variability in binaries on almost the shortest (scaled) time-scale on which we can study spectral variability in AGN. However, XRB systems offer the additional benefit of allowing us to study a much wider range of accretion rates within a single object, including transition across the \textit{\.{m}}$_{\rm crit}$ boundary, something which is impossible with an AGN.

In order, therefore, to measure the important $\Gamma$-\textit{\.{m}}$_{\rm E}$ diagnostic in a single black hole system, as it passes from the hard to the soft spectral state, across the \textit{\.{m}}$_{\rm crit}$ boundary, we select here fifteen observations from the \textit{Rossi X-ray Timing Explorer} (\textit{RXTE}) archive of the well known high mass X-ray binary (HMXB) system Cygnus X-1. This source was chosen because it has comprehensive archival coverage and is sufficiently bright to allow time-resolved X-ray spectroscopy on time-scales as short as $\rm 16~ms$. In order to study small changes in $\Gamma$ on these very short time-scales we measure $\Gamma$ from $\rm2~million$ separate observational sections (of time-scales between 16 and $\rm100~ms$, depending on flux level) and thus produce a $\Gamma$ lightcurve which we cross-correlate with the count rate lightcurve (used as an indicator of the accretion rate). Thus we can determine not only whether $\Gamma$ rises or falls with count rate but also examine whether or not changes in one of these parameters lags those in the other. In addition, we can measure the characteristic time-scale of the correlation (the width of the cross-correlation function) and examine how this evolves with accretion rate and hardness ratio over much longer time-scales.

\section{Observations}

All data used in this work have been selected and retrieved from the \textit{RXTE} archive, which has observed Cygnus X-1 frequently since 1996 and provides a vast repository of observations of this source. To simplify the reduction we have decided to use only those datasets that provide both Proportional Counter Array (PCA) standard-2 and generic binned array mode with at least $\rm16~ms$ time resolution. The standard-2 data provide good spectral resolution, with 47 channels available to cover the 3-$\rm20~keV$ energy range, but poor time resolution with data binned into intervals of 16~s. In contrast, the generic binned array modes provide excellent time resolution of typically 4-$\rm16~ms$ and the reduced spectral resolution helps to ensure a healthy number of counts per bin.

In total, 15 observations of Cygnus X-1 were retrieved from the \textit{RXTE} archive ranging in date from March 1996 until February 1998. These datasets, listed in Table \ref{tblObservations}, were selected to provide coverage of both the soft-state outburst in 1996 and the more steady hard state in which the source was found for the subsequent two years.

\section{Spectral fitting}

\subsection{Time-averaged fits to each complete observation}

In order to examine variability on time-scales of $\rm100~ms$ or less, we use the PCA generic binned array mode data. However, there are not enough counts, or spectral resolution, in each 16-$\rm100~ms$ section of these data to allow us to fit a complex model with all parameters free. We must therefore determine, from the lower time resolution, but higher spectral resolution, standard-2 data, a realistic model which can be applied to all the shorter time-scale spectral fits.

Initially, spectra were extracted from the standard-2 datasets for each full observation and fitted in \textsc{xspec} version 4.2 over the 3-$\rm20~keV$ energy range with a model consisting of absorption (\textsc{phabs}), a power-law component and its associated reflection (\textsc{pexrav}; \citealt{Magdziarz1995}), a Gaussian representing the iron K${\alpha}$ fluorescence line at $\rm\simnot6.4~keV$ (\textsc{gaussian}) and, if required, a disc black body component (\textsc{diskbb}). We fix the inclination angle of the \textsc{pexrav} component to $30^{\circ}$, and the fold energy, redshift, metal abundance and iron abundance to default values. Our choice of inclination angle is reasonably consistent with both the most recently published figure of ${\rm 27.1~\pm~0.8^{\circ}}$ \citep{Orosz2011} and also our own fits to the data, which vary by observation but tend to give best fit values slightly above $30^{\circ}$ in the hard state. However, when the inclination angle was fixed to a higher value of $53^{\circ}$ we found no discernible differences in the appearance of our count rate versus $\Gamma$ cross-correlation functions, and therefore suggest that the short time-scale, relative changes to $\Gamma$ such as we discuss in this paper are insensitive (or only weakly sensitive) to the choice of inclination angle.

In four of our fifteen observations (2, 3, 4 and 5) it was found that the addition of a disc black body component with an average disc temperature of $\rm0.47~keV$ significantly improved the fit. Hereafter, we refer to these four observations as being in the soft or intermediate states and the remaining eleven observations, that lacked an obvious thermal component, as being in the hard state. The soft/intermediate-state observations were also found to have broader average iron K${\alpha}$ lines than the hard states (one sigma Gaussian widths of $\rm1.31~keV$ and $\rm0.52~keV$ respectively) and stronger reflection scaling factors (1.66 and 0.46 respectively).

\begin{figure}
	\centering
	\includegraphics[width=86mm]{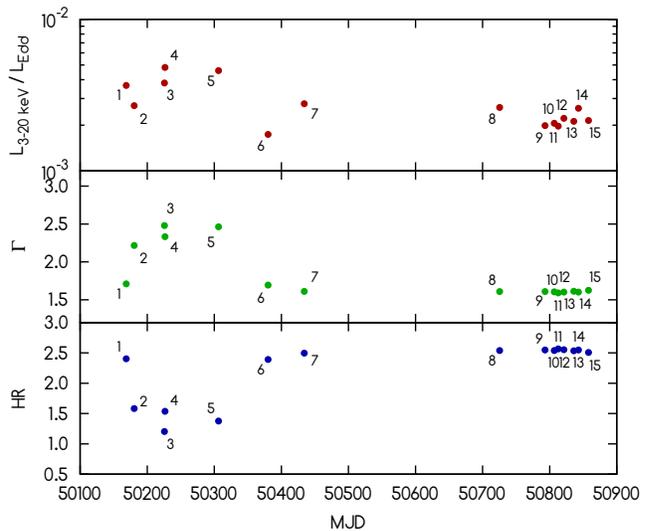}
	\caption{Plot shows the 3-$\rm20~keV$ X-ray luminosity to Eddington luminosity ratio, the photon index and hardness ratio ($L_{\rm5-13~keV}/L_{\rm3-5~keV}$) of Cygnus X-1 between March 1996 and February 1998. The source entered the soft state around March or April 1996. The numbers shown next to each datapoint identify the observation, and a summary of these is shown in Table \ref{tblObservations}.}
	\label{figCygX1LightCurve}
\end{figure}

The results of our spectral fitting are shown in Fig. \ref{figCygX1LightCurve}, which gives $\Gamma$, the hardness ratio ($L_{\rm5-13~keV}/L_{\rm3-5~keV}$) and the ratio of the 3-$\rm20~keV$ luminosity to the Eddington luminosity. The validity of the latter as a proxy for accretion rate is reduced somewhat by the potentially large differences in bolometric correction between the spectral states, and no attempt has been made here to estimate these. Furthermore, the sensitivity range of the PCA instrument does not provide coverage of the soft ($< \rm 2~keV$) or very hard ($> \rm 60~keV$) energy bands in which most of the power is emitted in the soft and hard states respectively \citep{Zdziarski2002,Churazov2001}, and the 3-$\rm20~keV$ luminosity may not therefore be an ideal choice for estimating accretion rate. The luminosity and Eddington limits were estimated using a distance of ${\rm 1.86^{+0.12}_{-0.11}~kpc}$ \citep{Reid2011} and a black hole mass of ${\rm 14.8~\pm~1.0~M_{\varodot}}$ \citep{Orosz2011}. The numeric labels correspond to the observation number as shown in Table \ref{tblObservations}. During the 1996 soft-state outburst the increase in $\Gamma$ and the reduction in the hardness ratio is clearly evident.

\subsection{Determining which model parameters to fix}

Although we now have suitable models for each of our observations, there are still not enough counts, or spectral resolution, on $\rm100~ms$ time-scales to allow all the model parameters to remain free during the fit. Fortunately, we expect that the power-law will be varying more on short time-scales than the other components so therefore adopt a strategy whereby the power-law is allowed to vary and all the other components remain fixed. We must be cautious, however, as although our models show that an iron line and a reflection component are important parts of the overall spectrum, they do not tell us whether the reflection scaling factor or the total flux in the reflected component remains constant over the whole observation. Similarly, we do not know if the flux in the iron line or the equivalent width of the iron line remains constant. Incorrectly fixing either one of these parameters, and particularly the reflection component, can lead to erroneous conclusions regarding the variation of $\Gamma$ with flux.

\begin{figure}
	\centering
	\includegraphics[width=86mm]{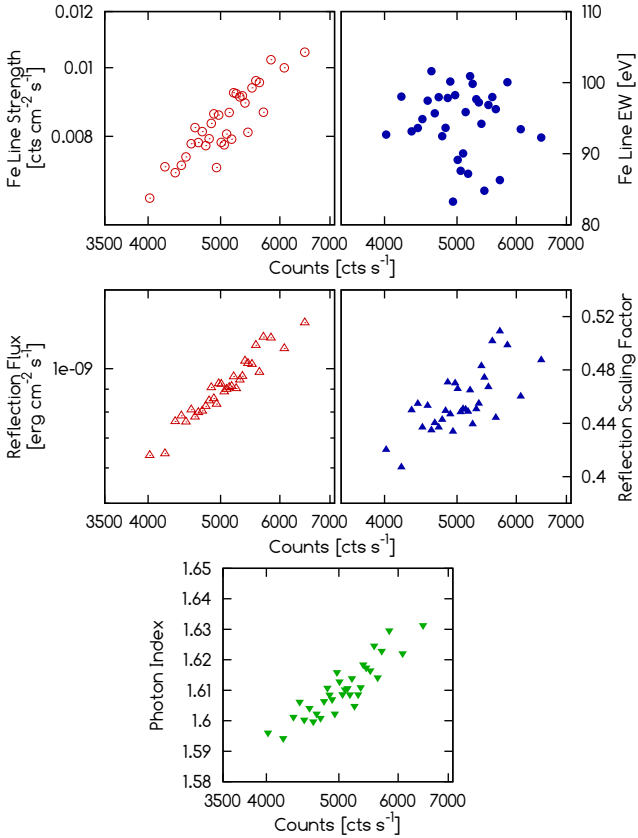}
	\caption{Relationship between the count rate and the properties of three of the spectral components for the hard-state observation 7 (1996 December 17). A total of 657 spectra, each $\rm16~s$ in length, were extracted from the standard-2 data and sorted into order of count rate. The $\rm16~s$ spectra were then summed to produce 33 new spectra each of length $\rm320~s$, which offer sufficiently high count rates to allow all the spectral components to vary during the fit. \textit{Top:} The strength of the iron K${\alpha}$ line at $\rm6.4~keV$, expressed in terms of its photon flux (left) and equivalent width (right). The photon flux scales with the count rate in the whole spectrum whereas the equivalent width remains roughly constant, suggesting that the latter should be fixed when fitting much shorter time-scale spectra. \textit{Centre:} The strength of the reflection component expressed in terms of its 3-$\rm20~keV$ flux (left) and its reflection scaling factor from the \textsc{pexrav} component (right). Both properties scale with the count rate (the latter to a lesser extent than the former) which could be indicative of either an increase in the optical depth of the disc or a reduction in the truncation radius at higher accretion rates. \protect\cite{Zdziarski1999} found a similar correlation between reflection and $\Gamma$ in a \textit{Ginga} sample of Seyferts and XRBs. \textit{Bottom:} The photon index of the power-law shows a strong correlation with the count rate.}
	\label{figStandard2Data}
\end{figure}

In order to determine which parameters in our models should be fixed, we extracted 657 spectra - each $\rm16~s$ in length - from the standard-2 data of the hard-state observation 7, sorted them into order of count rate and then summed them to produce 33 new spectra, each $\rm320~s$ in length. These data were fitted in \textsc{xspec} with \textit{all} components allowed to vary. The results are shown in Fig. \ref{figStandard2Data} and reveal that the equivalent width of the iron K$\alpha$ line remains roughly independent of the count rate (which has also been observed by \cite{Maccarone2002} for Cygnus X-1 and \cite{Nowak2002} for GX 339-4) and therefore fixing the equivalent width whilst fitting our $\rm100~ms$ spectra seems entirely reasonable. To do this, we replace the \textsc{gaussian} component in our models with the \textsc{gabs} component, and fix the line depth to a negative value in order to model an emission line. The reflection scaling factor, however, tends to increase slightly with the count rate (which is likely to be a short time-scale expression of the correlation between reflection and $\Gamma$ reported by \citealt{Zdziarski1999}), although not to the same extent as the total reflection flux. Therefore, we determine that fixing the reflection scaling factor seems more acceptable than fixing the whole reflection component.

These observations suggest an origin for the iron line, and most of the reflected component, in a close reprocessor such as the accretion disc. It is not clear why there is a small variation in the reflection scaling factor with flux. The positive sense of the correlation is opposite to that predicted by the light bending models designed to explain spectral variability \cite[e.g.][]{Miniutti2004} and, if interpreted within the context of light bending, would imply that the X-ray source becomes closer to the reflector as the overall luminosity goes up. Instead, this correlation may perhaps be attributed to a decrease in the disc truncation radius with increasing flux, which consequently increases the surface area of the reflector. Since we expect the variations in the fitted iron line equivalent width will be dominated by Poisson variations, rather than real ones, we hold the equivalent width fixed in order not to let random variations affect the other parameters.

\subsection{Fitting the spectra at $\rm100~ms$ time resolution}

Spectra were then extracted at $\rm100~ms$ time resolution from the binned array mode datasets of the observations listed in Table \ref{tblObservations}, and fitted in \textsc{xspec} over the 3-$\rm20~keV$ energy range. The absorption, reflection scaling factor, disc black body component and the equivalent width of the iron K${\alpha}$ line were fixed at the values determined from fits to each total observational dataset, and only the photon index and the normalisation of the power-law were allowed to vary. The spectra were extracted and fitted automatically by scripts that make use of the \textsc{heasarc} tools \textsc{saextrct} and \textsc{xspec}. For some of our observations it was also necessary to further rebin the data to improve the number of counts per bin, especially at energies above $\rm12~keV$.

\begin{table}
	\caption{A summary of the observations. Spectra were classified as either hard, soft or intermediate state depending upon the photon index and whether or not a disc black body component was required to fit the data in the 3-$\rm20~keV$ range.}
	\label{tblObservations}
	\begin{center}
	\begin{tabular}{@{}ccccc}
		\hline
		No. & Observation & Date & Spectral State & No.$^{a}$ \\
		& & & & Spectra \\
		\hline
		1 & 10238-01-08-00 & 26 Mar 1996 & Hard & 58,540 \\
		2 & 10238-01-04-00 & 07 Apr 1996 & Intermediate & 30,000 \\
		3 & 10412-01-02-00 & 22 May 1996 & Soft & 25,000 \\
		4 & 10412-01-01-00 & 23 May 1996 & Soft & 32,100 \\
		5 & 10412-01-05-00 & 11 Aug 1996 & Soft & 25,000 \\
		6 & 10241-01-01-00 & 24 Oct 1996 & Hard & 34,972 \\
		7 & 10236-01-01-02 & 17 Dec 1996 & Hard & 102,500 \\
		8 & 20175-01-02-00 & 04 Oct 1997 & Hard & 36,980 \\
		9 & 30157-01-01-00 & 11 Dec 1997 & Hard & 19,100 \\
		10 & 30157-01-03-00 & 24 Dec 1997 & Hard & 28,800 \\
		11 & 30157-01-04-00 & 30 Dec 1997 & Hard & 15,400 \\
		12 & 30157-01-05-00 & 08 Jan 1998 & Hard & 20,000 \\
		13 & 30157-01-07-00 & 23 Jan 1998 & Hard & 25,600 \\
		14 & 30157-01-08-00 & 30 Jan 1998 & Hard & 11,620 \\
		15 & 30157-01-10-00 & 13 Feb 1998 & Hard & 28,692 \\
		\hline
	\end{tabular} \\
	\end{center}
	\textsc{Notes}: $^{a}$ The number of $\rm100~ms$ spectra extracted from the observation. \\
\end{table}

\begin{figure}
	\centering
	\includegraphics[width=86mm]{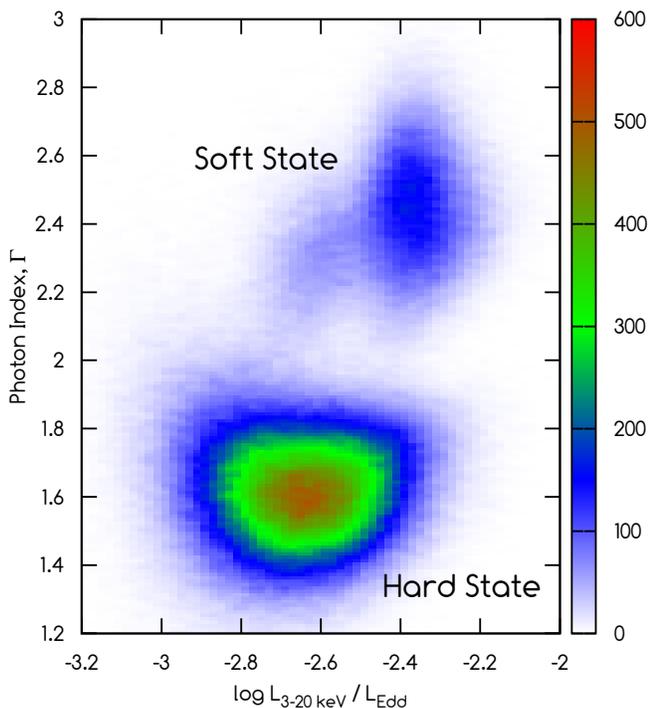}
	\caption{Density plot of $\rm\simnot500,000$ spectra, each $\rm100~ms$ in length, extracted and fitted from 11 hard-state and 4 soft/intermediate-state observations of Cygnus X-1. The plot shows the ratio of the X-ray luminosity to the Eddington luminosity ($L_{\rm 3-20~keV} / L_{\rm Edd}$) against the photon index ($\Gamma$). The hard and soft states are easy to distinguish with the latter forming a trail out towards the top of the plot. The scale on the right-hand-side refers to the number of spectra per pixel, where each pixel extends over a range $\Delta {\rm log}L=0.02$ on the x-axis and $\Delta \Gamma=0.01$ on the y-axis. It should be noted that the relative densities of the soft and hard states are biased by our choice of observations, and are not necessarily indicative of the amount of time the source spends in each state.} 
	\label{figDensityPlot}
\end{figure}

The range covered by these spectra is displayed in Fig. \ref{figDensityPlot}, which shows a density plot of approximately $\rm500,000$ spectra, each $\rm100~ms$ in length, that compares the X-ray luminosity (divided by the Eddington luminosity) with the photon index of the power-law component. The soft and hard states are easy to distinguish, with the former exhibiting a much higher average photon index (2.42, compared to 1.63) and higher luminosity ($L_{\rm 3-20~keV}$ was $\rm0.392\%$ of the Eddington limit, compared to $\rm0.224\%$). The cluster of points that appear intermediate between the soft and hard states are from observation 2 (7th April 1996), which relates to the start of the soft-state outburst and shows only a weak disc black body component in its X-ray spectrum.

\section{Cross-correlation of $\Gamma$ and count rate on 100 millisecond time-scales}

\subsection{The hard states}

\begin{figure}
	\centering
	\includegraphics[width=86mm]{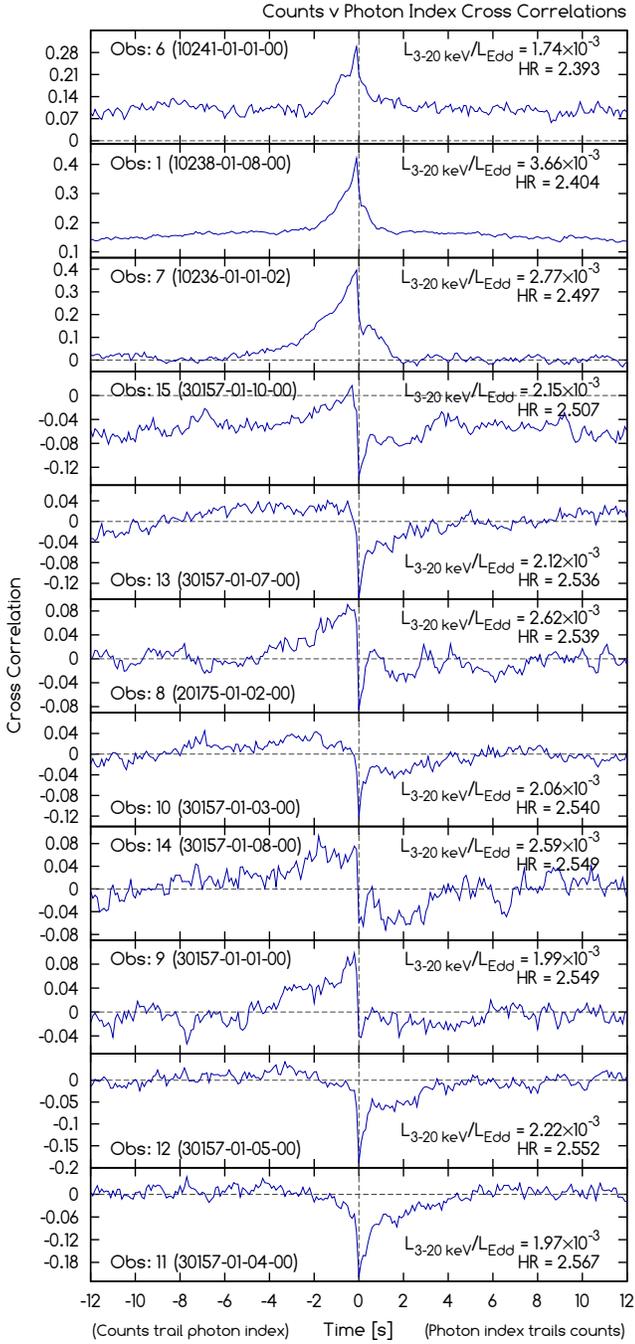}
	\caption{Cross-correlation functions between the photon index and the 3-$\rm20~keV$ source counts for 11 hard-state observations of Cygnus X-1. The observations are ordered by hardness ratio ($H.R.=L_{\rm 5-13~keV}/L_{\rm 3-5~keV}$), which is also shown on the right hand side of each panel along with the X-ray luminosity to Eddington luminosity ratio. The positive peaks imply that the source becomes softer when brighter and the negative peaks that the source becomes harder when brighter. The plots show a broad positive correlation and a narrower anti-correlation, the relative strength of which appears to be more strongly connected to the hardness ratio than the accretion rate. Although the strengths of some of the correlations in this plot are weak, they cannot be explained as the result of noise (see Appendix \ref{secNoise} for a discussion of this).}
	\label{figCrossCorrelations}
\end{figure}

In Fig. \ref{figCrossCorrelations} we show the cross-correlation functions (CCFs) between $\Gamma$ and the 3-$\rm20~keV$ count rate for all eleven of our hard-state observations at $\rm100~ms$ time resolution. We choose to use the discrete correlation function (DCF) method of \cite{Edelson1988} and normalise our results using the excess variance. The CCFs in this plot are ordered by hardness ratio ($H.R.=L_{\rm 5-13~keV}/L_{\rm 3-5~keV}$), which is also shown on the right hand side of each panel along with the accretion rate in Eddington units. We choose to use the source count rate instead of the flux in our CCFs as the latter, like $\Gamma$, is derived from our spectral fits and a degeneracy between these properties could produce erroneous results.

As each measurement of $\Gamma$ is subject to considerable statistical uncertainty, the peak values of some of the hard-state CCFs are not high. However, a large number of spectra contribute to each CCF (between $\rm\simnot10,000$ and $\rm\simnot100,000$ for each observation) and the results are none the less highly significant. A further discussion of noise in our data can be found in Appendix \ref{secNoise}.

The CCFs suggest the presence of positive and negative correlations, both peaking at around zero lag, and either of which can be missing from individual CCFs. Most of the plots reveal a strong lack of symmetry, with the positive correlation declining more slowly towards negative lags (source counts trail the photon index) and the anti-correlation always declining towards positive lags (photon index trails the source counts).

The most obvious conclusion to be drawn from examination of Fig. \ref{figCrossCorrelations} is that there is a steady transition from a positive to a negative correlation (i.e. an anti-correlation) as the hardness ratio increases. Some of the CCFs (particularly from observations 8, 9 and 14) show a greater degree of asymmetry than others, with a positive correlation being found at negative lags and a negative correlation at positive lags. We note that the changes in CCF shape occur over a quite limited range of hardness ratios. Examination of the data in the 2-$\rm3~keV$ range indicates that if the soft band was extended down to $\rm2~keV$, the hardness ratio range would expand. However as the response of the PCA below $\rm3~keV$ is both lower, and more uncertain, than at higher energies, we invoke the relatively standard procedure of only using data down to $\rm3~keV$.

The steady transition in the shape of the CCF is not seen so clearly if the observations are ordered by count rate. In particular we note that observation 6, which dates from a time when the source has just emerged from the soft state (24th October 1996), is both the softest and faintest observation and shows no sign of an anti-correlated component in its CCF. These observations suggest that the relative strength of the positive and negative correlations depends more strongly upon the hardness ratio than upon the accretion rate. We note, however, that given the large spectral changes that occur in this source \cite[e.g.][]{Zdziarski2002} the 3-$\rm20~keV$ count rate is far from being a perfect indicator of accretion rate, being particularly poor during transitions between states.

\subsection{The soft/intermediate states}

\begin{figure}
	\centering
	\includegraphics[width=86mm]{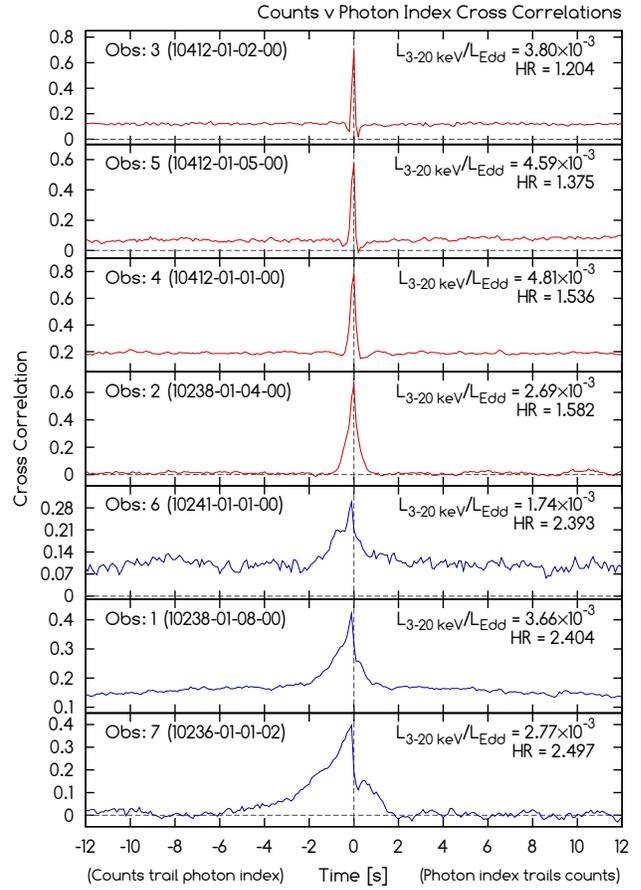}
	\caption{Cross-correlation functions between the photon index and the 3-$\rm20~keV$ source counts for four soft/intermediate-state observations of Cygnus X-1 (red lines) and, shown for comparison, the three hard-state observations with the lowest hardness ratios (blue lines). The CCFs are ordered by hardness ratio ($H.R.=L_{\rm 5-13~keV}/L_{\rm 3-5~keV}$), which is also shown on the right hand side of each panel along with the X-ray luminosity to Eddington luminosity ratio. All the CCFs show a single positive peak, centred at or near zero lag. Both the width of the peak and its asymmetry appear to increase with hardness ratio and there is no sign of the anti-correlation which is evident in many of the hard-state CCFs.}
	\label{figCrossCorrelationsSoft}
\end{figure}

The cross-correlation functions of photon index against source counts are shown in Fig. \ref{figCrossCorrelationsSoft} for our four soft/intermediate-state observations and, shown for comparison, the three hard-state observations with the softest hardness ratios. The soft-state CCFs show less complex profiles than the hard-state CCFs, with much stronger, narrower and more symmetric positive correlations. The anti-correlation that was evident in most of the hard-state CCFs is not obvious here.

\begin{figure}
	\centering
	\includegraphics[width=86mm]{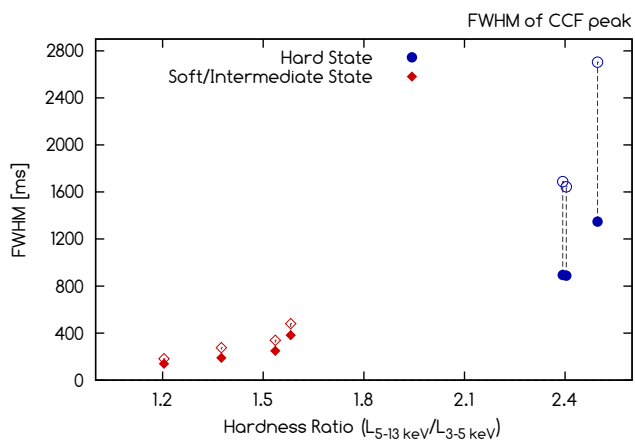}
	\caption{The widths of the peaks in the CCFs plotted against the hardness ratio for the four soft/intermediate states and the three hard-state observations with the lowest hardness ratios. Each of the peaks was found to be slightly asymmetric, with the width of the left-hand side slightly wider than that of the right. The symbols in the plot represent both the measured FWHM (filled symbols) and twice the width of the left-hand side (open symbols). In both cases the width is correlated with the hardness ratio, which possibly suggests a connection to the truncation radius of the accretion disc.}
	\label{figSoftHardnessVWidth}
\end{figure}

The widths of the peaks found in the soft/intermediate-state CCFs, along with those found in the hard-state observations 1, 6 and 7, have been measured (see Table \ref{tblPositivePeaks}) and the FWHM of the peaks are plotted against the hardness ratios ($L_{\rm 5-13~keV} / L_{\rm 3-5~keV}$) in Fig. \ref{figSoftHardnessVWidth}. Although this sample is limited in size, our data suggest that the width of the peak is correlated with the hardness ratio of the source but not the accretion rate.
 
 \begin{table*}
 	\caption{Properties of the four soft/intermediate-state CCFs, shown along with those of the three hard-state observations with the lowest hardness ratios. In this limited sample we find that the widths of the peaks in the CCFs show a stronger correlation with the hardness ratio than the accretion rate.}
 	\label{tblPositivePeaks}
 	\begin{center}
 	\begin{tabular}{@{}ccccccc}
 		\hline
 		No. & Observation & Date & $L_{\rm 3-20~keV} / {L_{\rm Edd}}^{a}$ & H.R.$^{b}$ & Width (FWHM)$^{c}$ & Width LHS$^{d}$ \\
 		& & & $\times10^{-3}$ & & [ms] & [ms] \\
 		\hline
		1 & 10238-01-08-00 & 26/03/1996 & 3.67 & 2.404 & 888 & 822 \\
 		2 & 10238-01-04-00 & 07/04/1996 & 2.69 & 1.582 & 382 & 240 \\
 		3 & 10412-01-02-00 & 22/05/1996 & 3.80 & 1.204 & 140 & 91 \\
 		4 & 10412-01-01-00 & 23/05/1996 & 4.81 & 1.536 & 249 & 169 \\
 		5 & 10412-01-05-00 & 11/08/1996 & 4.59 & 1.375 & 189 & 138 \\
 		6 & 10241-01-01-00 & 24/10/1996 & 1.74 & 2.393 & 895 & 844 \\
		7 & 10236-01-01-02 & 17/12/1996 & 2.82 & 2.497 & 1348$^{e}$ & 1352$^{f}$ \\
 		\hline
 	\end{tabular} \\
 	\end{center}
	\textsc{Notes}: $^{a}$ Ratio of X-ray luminosity to Eddington luminosity. $^{b}$ Hardness ratio ($L_{\rm 5-13~keV} / L_{\rm 3-5~keV}$). $^{c}$ FWHM of peak. $^{d}$ Width of the left-hand side of the peak, measured from the left-hand side of the half maximum up to the y-axis. $^{e,f}$ At half maximum the entire peak is to the left of the y-axis, and therefore the distance from the left-hand side to the y-axis is greater than the FWHM for this observation.
 \end{table*}

\section{Cross-correlation of $\Gamma$ and count rate on 16 millisecond time-scales}

\subsection{A closer look at the soft state, bright hard state and faint hard state}

All results discussed up to this point have been based upon $\rm100~ms$ time-resolution spectra, and separate cross-correlation functions have been generated for each observation. However, it would be useful to examine the differences between the soft states and the bright and faint hard states in a little more detail and, to do this, we approach the problem in two ways.

For the soft states and the bright hard states we choose to reduce the data again, but this time using $\rm16~ms$ time-resolved spectra. For the faint hard states we do not have sufficient counts to allow us to fit spectra on this time-scale, so we instead choose to combine the existing $\rm100~ms$ data from several observations and, from this combined data, produce a single cross-correlation function with improved signal-to-noise ratio.

\begin{figure}
	\centering
	\includegraphics[width=86mm]{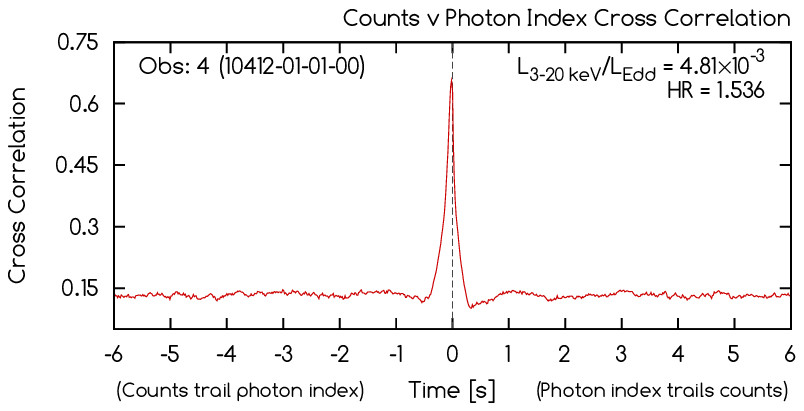}
	\includegraphics[width=86mm]{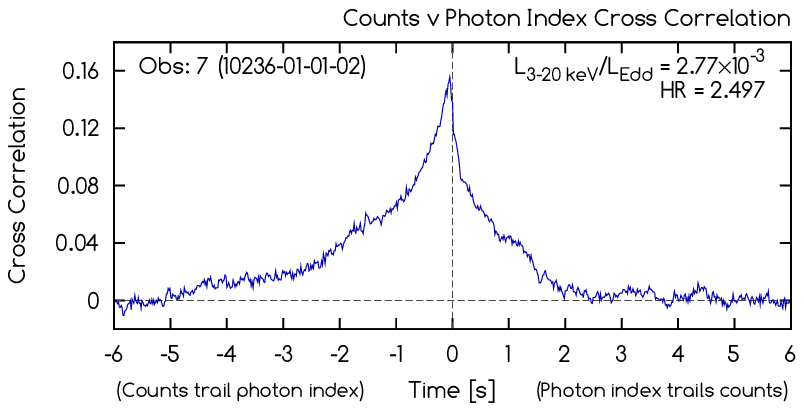}
	\includegraphics[width=86mm]{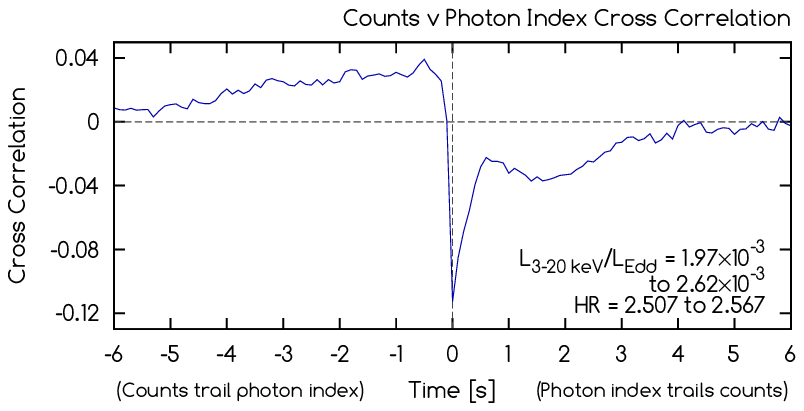}
	\caption{\textit{Top:} Cross-correlation function (CCF) between the photon index and the 3-$\rm20~keV$ source counts for the soft-state observation 4 (1996 May 23), shown with $\rm16~ms$ time resolution. The peak in the CCF is smooth and featureless, even on time-scales as short as these, but is slightly asymmetric. \textit{Centre:} CCF between the photon index and the 3-$\rm20~keV$ source counts for the bright hard-state observation 7 (1996 December 17), shown with $\rm16~ms$ time resolution. The peak in the CCF is non-symmetric with a gradual decline towards negative time-lags and a steeper decline towards positive time-lags. \textit{Bottom:} CCF between the photon index and the 3-$\rm20~keV$ source counts, produced by combining the data from 8 low accretion rate/high hardness ratio observations of Cygnus X-1. The plot reveals a narrow, non-symmetric anti-correlation in which the photon index trails the counts by a period of less than one second. In addition to this peak, we find the two properties are weakly correlated at negative time-lags and weakly anti-correlated at positive time lags.}
	\label{figCrossCorrelationsHighRes}
\end{figure}

Fig. \ref{figCrossCorrelationsHighRes} (top) shows the CCF for the soft-state observation 4 (1996 May 23) using $\rm16~ms$ time-resolved spectra. This plot was generated from a total of $\rm199,989$ spectra, with an average of 147 counts in the 3-$\rm20~keV$ range and a mean photon index of 2.30. We see that the peak in the CCF is smooth and featureless even on time-scales as short as these, but we also find a very slight asymmetry that was not clear at $\rm100~ms$ time resolution.

Fig. \ref{figCrossCorrelationsHighRes} (centre) shows the CCF for the bright hard-state observation 7 (1996 December 17), also using $\rm16~ms$ time-resolved spectra. This plot was generated from a total of $\rm361,868$ spectra, with an average of 66 counts in the 3-$\rm20~keV$ range and a mean photon index of 1.48. The model is the same as that which was used for the soft-state observation, except the disc black body component was not required. Although the peak is found close to zero lag, it is not quite symmetric in appearance and is instead skewed towards negative time lags.

Fig. \ref{figCrossCorrelationsHighRes} (bottom) shows the CCF for the combined faint hard-state observations (8 to 15), using $\rm100~ms$ time-resolution spectra. The narrow anti-correlation is clearly evident and is found mostly at positive lags, while the positive correlation is either missing or has been significantly broadened.

The peaks found in the CCFs of Figs. \ref{figCrossCorrelations} and \ref{figCrossCorrelationsSoft} show a degree of asymmetry that tends to increase with the hardness ratio. We find that the CCFs that show a positive correlation between source counts and $\Gamma$ often have peaks that are skewed towards negative lags and those showing an anti-correlation are always skewed towards positive lags. A possible explanation for this could be that the positive and negative correlation peaks are actually symmetric in each of our observations, but as the strength of the correlation decreases the presence of another, non-symmetric, component begins to become more apparent, resulting in a peak that is broader on one side of the y-axis than it is on the other.

\begin{figure}
	\centering
	\includegraphics[width=86mm]{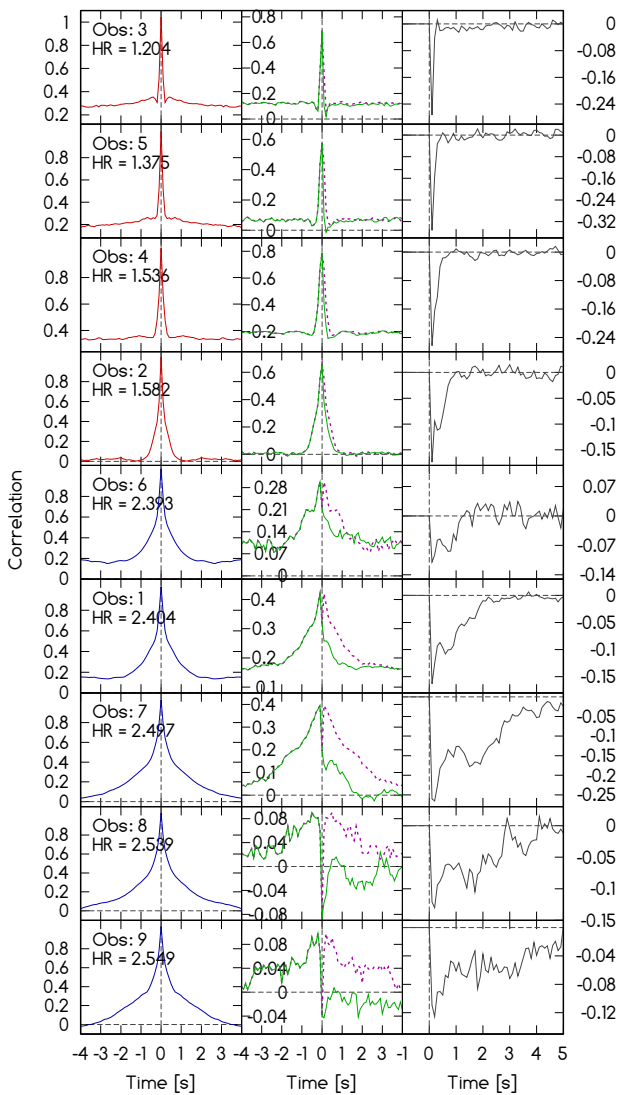}
	\caption{\textit{Left panel:} Autocorrelation functions (ACFs) generated from the source counts of four soft/intermediate-state and five hard-state observations of Cygnus X-1. The width of the peak increases as the hardness ratio increases, as shown in Fig. \ref{figSoftHardnessVWidth}. This is consistent with the findings of \protect\cite{Pottschmidt2003}, who reported that characteristic features of the power spectral density (PSD) shift towards higher frequencies as the spectra soften. \textit{Centre panel:} Cross-correlation functions between the source counts and the photon index (solid green line). We note that the width of the peak is always greater on the left-hand side of the y-axis than the right-hand side and suggest that this may be caused by the presence of at least three components in the CCFs - a positive symmetric correlation, a negative symmetric correlation and an asymmetric component. Also shown on the same plots is a reflection in the y-axis of the left hand side of the CCFs (dotted purple line). \textit{Right panel:} The difference between the right hand side of the CCF and the reflection of the left hand side in the y-axis (green line minus purple line). Many of these plots appear similar to the CCFs produced from the faintest hard-state observations (see Fig. \ref{figCrossCorrelations}), which lacked a strong positive correlation, and may therefore suggest that the asymmetric component that was clearly evident in the hard-state observations may also be present in the soft-state observations.}
	\label{figACF}
\end{figure}

In Fig. \ref{figACF} (left panel) we plot the autocorrelation functions (ACFs) of the source counts for four soft/intermediate-state and five hard-state observations and find that the widths of the peaks, especially those generated from the soft-state observations, are reasonably consistent with the widths of the peaks in the source counts/photon index CCFs (Fig. \ref{figACF}, centre panel). The ACF widths tend to increase with the hardness ratio, which is consistent with reports that the power spectral density (PSD) shifts towards higher frequencies as the spectra soften \cite[see][and references therein]{Pottschmidt2003}. We then compare the right-hand side of the CCFs with the reflection of the left-hand side (shown as a dotted purple line in Fig. \ref{figACF}, centre panel) and, in the rightmost panel of the same figure, plot the difference between the two. The result is not unlike the anti-correlation peak found in some of the hard-state CCFs (Fig. \ref{figCrossCorrelations}) and raises the possibility that this highly asymmetric component may be present in \textit{all} of our CCFs, including those from the soft-state observations.

\subsection{A lag between the soft and hard counts}

\begin{figure}
	\centering
	\includegraphics[width=86mm]{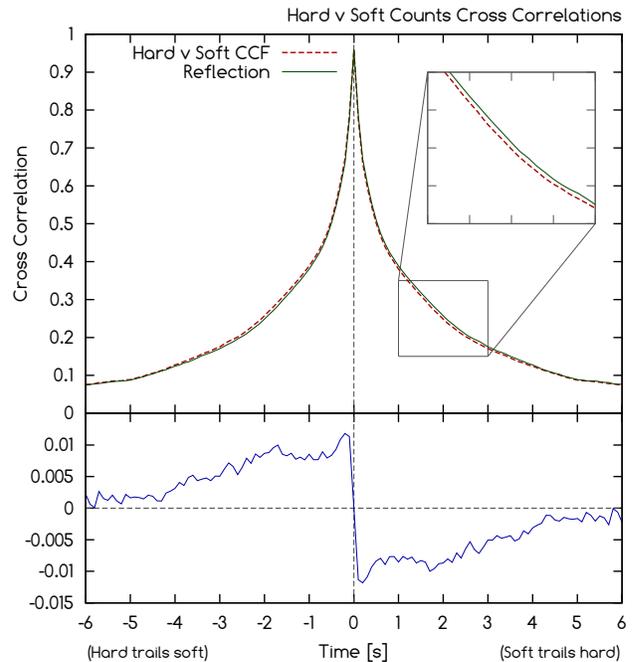}
	\caption{Cross-correlation function between the hard (5-$\rm13~keV$) count rate and the soft (3-$\rm5~keV$) count rate (dashed red line) for the combined lightcurve from 8 hard-state observations. The function is almost symmetric, but when plotted alongside its own reflection in the y-axis (solid green line) the slight asymmetry reveals that the hard count rate, or some component of it, slightly lags the soft count rate. The bottom panel emphasises this point by showing the difference between these two functions. In Section \ref{lblDiscussion} we consider the effect that this lag may have on our measurements of $\Gamma$, and suggest that it may be the most likely explanation for the asymmetric shape of our $\Gamma$ versus counts CCFs.}
	\label{figLags}
\end{figure}

Further analysis of the combined $\rm100~ms$ data from the eight hard-state observations reveals a small asymmetry in the CCF (Fig. \ref{figLags}, dashed red line) of hard counts (5-$\rm13~keV$) versus soft counts (3-$\rm5~keV$), which can only be produced if a small component of the hard count rate lags the changes in the soft count rate. Although the peak in the CCF is almost symmetric, when plotted alongside its own reflection in the y-axis (i.e. the CCF of soft counts against hard counts, solid green line) the slight asymmetry becomes apparent, and can be further emphasised by subtracting one CCF from the other (bottom panel, solid blue line). This asymmetry in the CCF of Cygnus X-1 was first observed by \cite{Priedhorsky1979} and \cite{Nolan1981}, and later re-examined in more detail by \cite{Maccarone2000}, who ruled out light travel delays in the corona as an explanation.

In Section \ref{lblDiscussion} we consider the effect that this lag may have on our measurements of $\Gamma$, and suggest that it may be the most likely explanation for the asymmetric shape of our $\Gamma$ versus counts CCFs.

\section{A comparison with GX 339-4}

\begin{figure}
	\centering
	\includegraphics[width=86mm]{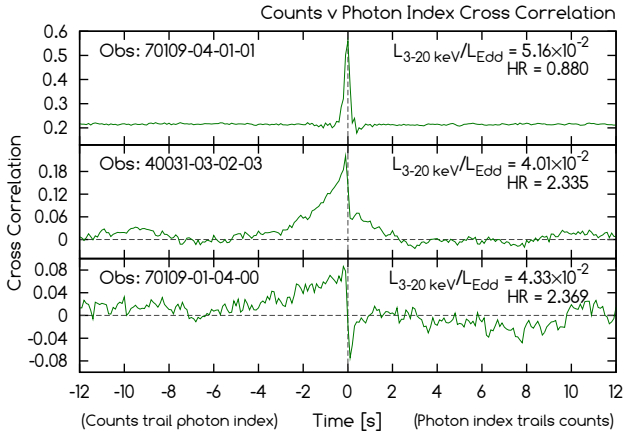}
	\caption{Cross-correlation functions between the photon index and the 3-$\rm20~keV$ source counts for the low-mass X-ray binary (LMXB) system GX 339-4. The panels show three observations from the \textit{RXTE} archive, dating from April/May 2002, and ordered by hardness ratio ($L_{\rm 5-13~keV} / L_{\rm 3-5~keV}$). The results show similar behaviour to that found in the soft-state and bright hard-state observations of Cygnus X-1 (Fig. \ref{figCrossCorrelations}), with the shape of the CCFs reminiscent of observations 1, 8 and 10. The anti-correlation found in the faint hard state of the Cygnus X-1 CCFs is also present here in the observation with the greatest hardness ratio.}
	\label{figCrossCorrelationsGX339}
\end{figure}

In order to investigate whether the results discussed in this paper are unique to Cygnus X-1 or are also found in other Galactic black holes we have extended our analysis to include the low mass X-ray binary (LMXB) system GX 339-4. In the quiescent state, this source is generally much too faint to allow spectra to be fitted on $\rm100~ms$ time-scales and we are therefore restricted to using only those observations that were made at times when GX 339-4 was experiencing an outburst. However, although the source is sufficiently bright at these times, we often find that most of the photons are coming from a thermal component and not from the power-law, which can still be difficult to fit on short time-scales. We have attempted to resolve this problem by selecting datasets from times when the source was ascending towards its maximum brightness (around April and May 2002, during which time the source could be found first in a hard state, and then later a soft state), and neglecting the softer state observations which followed in subsequent months. For a discussion of GX 339-4 during this outburst see \cite{Belloni2005}. The cross-correlation functions produced from these data are shown in Fig. \ref{figCrossCorrelationsGX339} and reveal similar behaviour to that found in the CCFs of Cygnus X-1, namely a positive peak that becomes wider and more asymmetric as the hardness ratio increases and, in the observation with the greatest hardness ratio, the emergence of a narrow anti-correlation skewed towards positive lags.

\section{Discussion}
\label{lblDiscussion}

\subsection{Summary of the main observational results}
\label{lblSummaryObservationalResults}

$\bullet$ In the hardest states of Cygnus X-1 which we have examined, the photon index, $\Gamma$, is anti-correlated with the count rate. Where a positive time lag is defined as $\Gamma$ lagging behind the count rate, there is little correlation at negative lag. However, a strong anti-correlation occurs very sharply within $\rm16~ms$ of zero lag and then decreases over the following few seconds. The strongest anti-correlation has a correlation coefficient (\mbox{$\simnot$-0.2}) which is relatively weak when compared to the strength of the positive correlations found in the soft state.

$\bullet$ In the softest states which we have examined, $\Gamma$ increases as the count rate increases (i.e. there is a positive correlation between the two parameters). The CCF is nearly symmetric about zero lag and the peak correlation coefficient is quite high ($>0.75$) - see Fig. \ref{figCrossCorrelationsSoft}.

$\bullet$ When the data from the three softest of the hard/intermediate state observations are excluded (i.e. observations 1, 6 and 7), a CCF generated from the remaining hard-state data (Fig. \ref{figCrossCorrelationsHighRes}, bottom panel) shows a weak, broad, positive correlation at negative lags and a weak, broad, negative correlation at positive lags, together with a sharp negative peak in the CCF near zero lag. The weak and broad components of the CCF are remarkably close in shape to that which was produced when we examined the difference between the hard count rate versus soft count rate CCF and its reflection about zero lag (Fig. \ref{figLags}, bottom panel).

$\bullet$ There is an almost monotonic transition with varying hardness ratio between the shapes of the CCF in the softest and hardest states with the CCFs in the intermediate hardness states being simply replicated as a sum of the two extreme CCFs, each added in proportion with hardness ratio.

$\bullet$ The full width of the CCFs in the soft state, together with twice the half-width at negative lags in the harder states, correlates well with hardness ratio, as does the full width of the ACF in all states.

$\bullet$ Although we have not yet investigated the behaviour of other XRBs in as much detail, we have presented some early indications that the CCFs of GX 339-4 exhibit similar features to those of Cygnus X-1, suggesting that this behaviour may be common to all black hole XRBs.

\subsection{The hard and soft-state emission components}

The CCFs provide clear evidence for two distinct components contributing to the X-ray emission. The soft-state CCFs are the simplest to explain. They show the standard behaviour expected when an increased seed photon flux, perhaps resulting from an increase in the accretion rate through the disc, is sufficiently large that it cools the Comptonising corona and leads to both an increased X-ray flux and also a softer spectrum \citep{Laurent2011, Qiao2013}. Thus energy is released primarily through the production of seed photons rather than in heating the corona. Seed photon production is most efficient from the optically thick part of the accretion disc and the small width of the CCF is consistent with a change in the seed photon flux on a viscous time-scale associated with an accretion disc at small gravitational radius. The very direct link between seed photon flux and coronal cooling leads to a relatively high peak CCF value and a lag close to zero. The decreasing width of the CCF with decreasing hardness is consistent with the inward motion of the inner part of the disc with increasing accretion rate.

The hard-state CCFs are more complex. As we explained above (Section \ref{lblSummaryObservationalResults}), they appear to consist of a broad anti-symmetric component and a narrow symmetric component (see Fig. \ref{figCrossCorrelationsHighRes} - bottom panel). The narrow width of the symmetric component implies that the origin of the flux change and of the spectral index change for this component have to be very closely coupled and in the same physical location. Cyclo-synchrotron self-Compton emission, where the cyclo-synchrotron seed photons are scattered by their very nearby parent electrons, naturally provides this component to the CCF, and should dominate in the highest density, inner regions of the corona or RIAF where the locally produced cyclo-synchrotron photons may form the bulk of the seed photons. Enhanced cyclo-synchrotron radiation implies that more high energy electrons have been injected, i.e. energy has been put into the Comptonising electrons of the corona rather than into the seed photons.
 
\begin{figure}
	\centering
	\includegraphics[width=84mm]{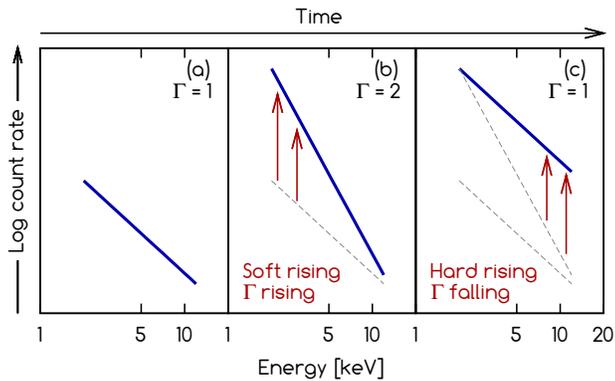}
	\caption{An illustration of an X-ray spectrum with a single power-law component (panel \textit{a}), demonstrating how a lag between the soft count rate and the hard count rate may temporarily affect the measurement of $\Gamma$ and potentially influence the CCF of count rate versus $\Gamma$. An increase in count rate is initially observed more strongly at soft energies, which causes the spectrum to temporarily soften (panel \textit{b}). At some later time the hard energy count rate will also start to rise and the spectrum will now begin to harden (panel \textit{c}). Together, one might expect this softening and hardening of the spectrum to produce a positive correlation between $\Gamma$ and the total count rate at negative lags ($\Gamma$ leads count rate) and an anti-correlation at positive lags (count rate leads $\Gamma$). Observationally, the hard-state observations provide some indication that this may be the case (see Fig. \ref{figCrossCorrelationsHighRes}, bottom, and Fig. \ref{figCrossCorrelations}, observations 10, 13 and 14).}
	\label{figCartoonSpectrum}
\end{figure}

To produce the broad anti-symmetric component we require that the soft part of the spectrum rises first, leading to a steepening of the spectrum with count rate. We then require that, some time later, the count rate in the higher energy parts of the spectrum rises, leading to a hardening of the spectrum with count rate. Thus, as the count rate rises, a steepening of the spectrum ($\Gamma$ is positively correlated with the count rate) precedes a hardening of the spectrum ($\Gamma$ is anti-correlated with the count rate) and introduces the weak asymmetric component we see in our CCFs. This scenario, which requires that some component of the hard flux lags the soft, agrees entirely with the hard versus soft band CCF (Fig. \ref{figLags}) and is illustrated in Fig. \ref{figCartoonSpectrum}. Such spectral changes can be explained physically if there is an extended emission region with a radial gradient in the energy or temperature of the Comptonising electrons such that the higher energy electrons are mostly closer to the black hole than are the lower energy electrons. Thus the emission weighted centroid of the high energy photons would be closer to the black hole than that of the lower energy photons. An inwardly moving perturbation, perhaps in accretion rate, would then boost the low energy photon emission rate before that of the higher energy photons. This scenario is entirely consistent with what is now becoming the standard paradigm to explain most aspects of X-ray spectral variability, i.e. accretion rate perturbations, produced probably in the accretion disc \citep{Lyubarskii1997}, propagating inwards and affecting the X-ray emission from a radially temperature-stratified emission region \cite[e.g.][]{Kotov2001,Churazov2001,Arevalo2006,Arevalo2008}.

For the above propagating fluctuation model to work we require that the main X-ray production mechanism at intermediate hardness is not self-Compton scattering of locally produced cyclo-synchrotron photons, which would only produce a narrow anticorrelation. Instead, we require that, outside of the densest central region of the corona, the seed photons will be dominated by a mixture of thermal disc seed photons and a significant component of cyclo-synchrotron photons produced in more distant regions, although there will also be some locally produced cyclo-synchrotron seed photons. Thus the factor which mainly drives flux variability in the outer regions will be enhancement of the scattering electron population rather than seed photon enhancement and so lags will indeed be expected between soft and hard photons as the perturbations pass through the energy stratified scattering region.

\begin{figure}
	\centering
	\includegraphics[width=84mm]{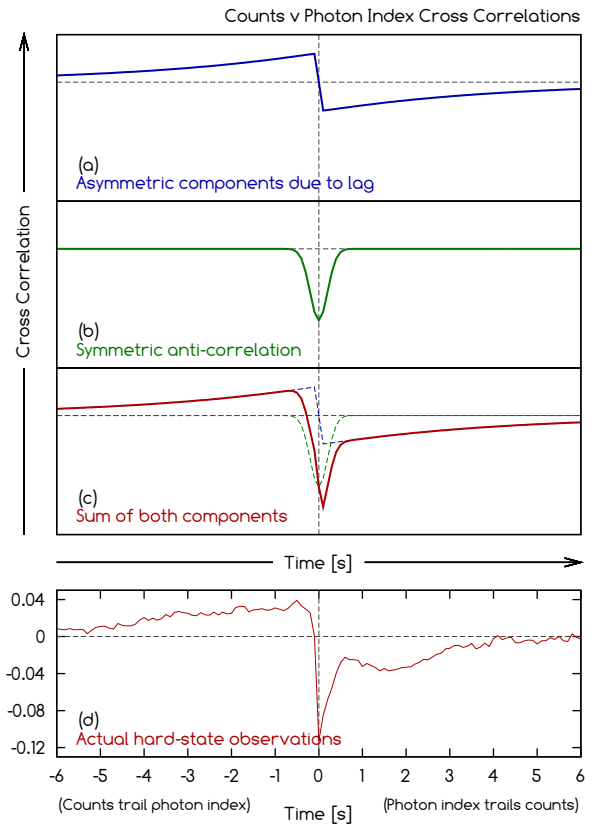}
	\caption{An illustration of a CCF between count rate and $\Gamma$, demonstrating how a lag between the soft count rate and the hard count rate may be responsible for the asymmetric shape of the hard-state CCFs. \textit{Panel a:} The highly asymmetric components introduced into the CCF by small changes in $\Gamma$ as variations in the hard count rate lag those in the soft count rate (see Fig. \ref{figCartoonSpectrum}). \textit{Panel b:} The narrow anti-correlation which we tentatively associate with Comptonisation from cyclo-synchrotron photons in the hot inner flow. \textit{Panel c:} The sum of the components in panels a and b. \textit{Panel d:} The CCF generated from eight hard-state observations, showing a positive correlation at negative lags, a negative correlation at positive lags and a narrow, asymmetric anti-correlation.}
	\label{figCartoonCCF}
\end{figure}

Fig. \ref{figCartoonCCF} (panel \textit{a}) shows an illustration of how an asymmetric component, generated purely from our simple model of an energy/temperature gradient and propagation lag rather than any genuine change in $\Gamma$, might appear on the CCF of count rate versus $\Gamma$. When added to a narrower, symmetric anti-correlation (panel \textit{b}) we find that the result is a skewed anti-correlation (panel \textit{c}) that has at least some potential for reproducing the overall shape of the CCF generated from the eight hard-state observations (panel \textit{d}).

\section{Acknowledgements}

We gratefully thank Chris Done for the very helpful discussion and also the referee, Andrzej Zdziarski, for the useful comments. This research has made use of data obtained through the High Energy Astrophysics Science Archive Research Center Online Service, provided by the NASA/Goddard Space Flight Center. CJS acknowledges support from a STFC studentship.

\bibliographystyle{mn_antonysmith}
\bibliography{references}

\appendix
\section{The effect of noise on our results}
\label{secNoise}

\begin{figure}
	\centering
	\includegraphics[width=84mm]{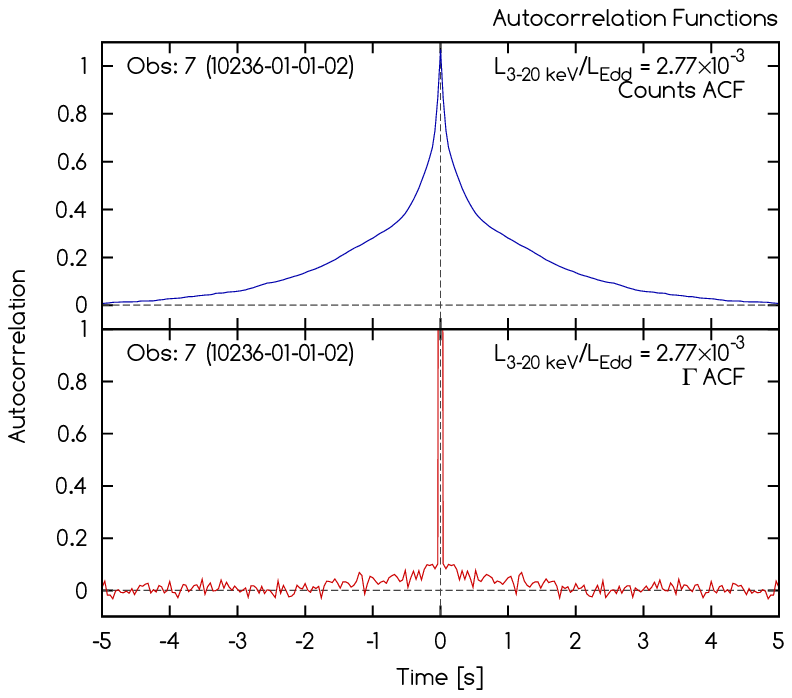}
	\caption{Autocorrelation functions (ACFs) for the hard-state observation 7, showing source counts (top) and $\Gamma$ (bottom). The extremely narrow peak in the $\Gamma$ ACF shows that these data are heavily affected by noise, which was already suspected from the large amount of scatter that was evident in Fig. \ref{figDensityPlot}. This does not mean, however, that $\Gamma$ should be entirely uncorrelated with the source counts nor that this noise is responsible for generating the peaks found in the CCFs.}
	\label{figACF40ms}
\end{figure}

In the hard-state CCFs shown in Figure \ref{figCrossCorrelations}, the strongest anti-correlation has a correlation coefficient of approximately \mbox{-0.2}. Such a weak correlation may be the result of noise in the data and it is therefore worth exercising a little caution. Fig. \ref{figACF40ms} shows the autocorrelation functions (ACFs) of the source counts and $\Gamma$ for the hard-state observation 7, shown at $\rm 40~ms$ time resolution. The ACF of a lightcurve consisting of just white noise would have a form similar to a delta function and it is clear from the narrow peak that $\Gamma$ is heavily affected by noise. This does not mean, however, that $\Gamma$ is entirely uncorrelated with the source counts and, since a large number of spectra (between $\simnot10,000$ and $\simnot100,000$ for each observation) contribute to each CCF, we believe that the results are still highly significant.

\begin{figure}
	\centering
	\includegraphics[width=84mm]{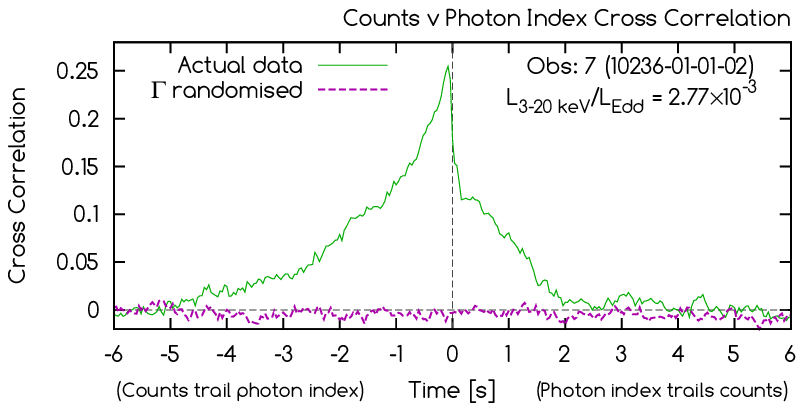}
	\caption{A comparison between the CCF for observation 7 and a new CCF, generated by completely randomising the photon index data while leaving the source counts unaltered. The new CCF shows small amplitude variations that are at least an order of magnitude weaker than the asymmetric peak we see in the real data CCF.}
	\label{figNoise}
\end{figure}

In order to test this assertion we have generated a new CCF by completely randomising the photon index data while leaving the source counts unaltered (Fig. \ref{figNoise}). This new CCF shows small amplitude variations that are more than an order of magnitude weaker than the peak we see in the CCF of the real data.

\end{document}